\documentclass{PoS}
\pdfoutput=1
\usepackage[pdftex]{graphicx}
\usepackage{subfig}
\usepackage{float}
\usepackage{url}
\usepackage{braket}
\usepackage{amsmath}
\usepackage{amssymb}
\usepackage{esint}
\usepackage{dsfont}

\graphicspath{{figures/}}

\newcommand{\Tr}{\mbox{\rm Tr\,}}

\title{Lattice SU(2) on GPU's}

\ShortTitle{Lattice SU(2) on GPU's}

\author{Nuno Cardoso \\
CFTP, Instituto Superior T\'{e}cnico\\
E-mail: \email{nunocardoso@cftp.ist.utl.pt}
}

\author{Pedro Bicudo \\
CFTP, Instituto Superior T\'{e}cnico\\
E-mail: \email{bicudo@ist.utl.pt}
}

\abstract{
    We discuss the CUDA approach to the simulation of pure gauge Lattice SU(2).

   CUDA is a hardware and software architecture developed by NVIDIA for computing on the GPU.
   We present an analysis and performance comparison between the GPU and CPU with single precision.
   Analysis with single and multiple GPU's, using CUDA and OPENMP, are also presented.
   In order to obtain a high performance, the code must be optimized for the GPU  architecture, i.e., an  implementation that exploits the memory hierarchy of the CUDA programming model.

   Using GPU texture memory and minimizing the data transfers between CPU and GPU, we achieve a speedup of $200\times$ using 2 NVIDIA 295 GTX cards relative to a serial CPU, which demonstrates that GPU's can serve as an efficient platform for scientific computing.
   With multi-GPU's we are able, in one day computation, to generate 1 000 000 gauge configurations in a $48^4$ lattice with $\beta=6.0$ and calculate the mean average plaquette.

   We present results for the mean average plaquette in several lattice sizes for different $\beta$. Finally we present results for the mean average Polyakov loop at finite temperature.
}

\FullConference{The XXVIII International Symposium on Lattice Field Theory - LAT2010\\
		 July 14-19 2010\\
		 Villasimius, Sadinia, Italy}

\begin{document}

\section{Introduction}

In the last years, the developments in computer technology made graphics processing units (GPU's) available to everyone as a powerful coprocessor in numerical computations.
Currently  many-core GPU's such as NVIDIA GTX and Tesla GPU's can contain more than 500 processor cores on one chip in the recent architecture, Fermi architecture.
However, GPU's are known to be hard to program, since coalescing of parallel memory is a critical requirement for achieve the maximum performance.

In this paper, we explore the limits of GPU computing in generating SU(2) pure lattice configurations using the heat bath method and performing some measurements in this configurations like the mean average plaquette and the Polyakov loop in single precision. We test the performance using 1 GPU, 2 GPU's and 4 GPU's. We use two NVIDIA 295 GTX cards, each card has two GPU's inside.

This paper is divided in 5 sections. In section 2, we present a little description in how to generate lattice SU(2) configurations, in section 3 we present the Cuda programming model, in section 4 we present the GPU performance over one CPU core as well as results for the mean average plaquette and Polyakov loop for different $\beta$ and lattice sizes. Finally, in section 5 we present the conclusions.

\section{Lattice SU(2)}

A matrix $\textrm{U}\in\text{SU(2)}$ can be written as:
$$\textrm{U}=a_0\, \mathds{1} + i\textrm{a}\cdot \sigma$$
with
$$a^2=a_0^2+\textrm{a}^2=1$$
This condition defines the unitary hyper-sphere surface $S^3$,
$$\textrm{dU}=\frac{1}{2\pi^2}\delta\left(a^2-1\right)\textrm{d}^4a$$
and
$$\Tr \textrm{U}=2a_0,\quad \quad\textrm{U}\textrm{U}^{\dagger}=\mathds{1},\quad \quad \det \textrm{U} = 1$$

To generate SU(2) pure gauge configurations, \cite{Creutz,Creutz1}, we need to calculate the staple, $V$ the sum of the six products of neighboring links that interact with a particular link, in the heat bath method. The distribution to be generated for every single link is,
       $$\textrm{d}P(\textrm{U})\propto \exp \left[\frac{1}{2} \beta \Tr(UV)\right]$$
Using a useful property of SU(2) elements, saying that any sum of them is proportional to another SU(2) element
        $$\tilde{\textrm{U}}=\frac{V}{\sqrt{\det V}}=\frac{V}{k}$$
 where $\tilde{\textrm{U}}$ is another SU(2) element and the invariance of the group measure
        $$\textrm{d}P\left(\textrm{U}\tilde{\textrm{U}}^{-1}\right) \propto \exp\left[\frac{1}{2}\beta k \Tr \textrm{U}\right]\textrm{dU} = \exp\left[\beta k a_0\right] \frac{1}{2\pi^2}\delta\left(a^2-1\right)\textrm{d}^4 a$$  
We generate a random number for $a_0\in [-1,1]$ with distribution
        $$P\left(a_0\right)\sim \sqrt{1-a_0^2} \exp\left(\beta k a_0\right)$$
once the $a$'s are obtained in this way, the new link is updated,
       $$\textrm{U}'=\textrm{U}\tilde{\textrm{U}}^{-1}$$

\section{Cuda Programming Model}

NVIDIA Compute Unified Device Architecture (CUDA), \cite{cuda, cudaguide} is a general purpose parallel computing architecture with a novel parallel programming model and instruction set architecture.
C for CUDA exposes the CUDA programming model as an abstraction of GPU parallel architecture using
a minimal set of extensions to the C language by allowing programmers to define C functions, called kernels.

In general, unlike CPU, a GPU has more transistors dedicated to data processing than to data caching and flow control.
A basic building block of NVIDIA GPUs is a multiprocessor with 8 cores, up to 16384 32-bit registers, 16KB memory shared between 1024 co-resident threads (a multiprocessor executes a block of up to 16 warps, comprising of up to 32 threads, simultaneously).
Conditional execution is possible, but groups of 32 threads (a thread warp) must execute the same instruction in SIMD fashion.
With up to 240 cores (30 multiprocessors) and memory bandwidth up to 102 GBps, the latest generation of GPUs offers extremely cost-effective computational power not only for visualization but also for general purpose scientific computations.

NVIDIA GPU memory model is highly hierarchical and there exist per-thread local memory, per-thread-block shared memory and device memory which aggregates global, constant and texture memory allocated to a grid, an array of thread blocks.
A thread executes a kernel, GPU program, and communicates with threads in the same thread block via high-bandwidth low-latency shared memory.
Generally, optimizing the performance of CUDA applications could involve optimizing data access patterns to these various memory spaces.
Each of the memory space has certain performance characteristics and constraints.
To have an efficient implementation is necessary to consider carefully CUDA memory spaces, specifically, local and global memories which are not cached and have high access latencies.
The CPU, host, can read and write global memory but not to shared memory.
    
A kernel is a function called from CPU that runs on the GPU. A CUDA kernel is executed by an array of threads, all threads run the same code and each thread has an ID that it uses to compute memory addresses and make control decisions.
A kernel launches a grid of thread blocks. Threads within a block cooperate via shared memory, however threads in different blocks cannot cooperate.

\section{Results}

We implement our code in CUDA language to run in one GPU or in several GPU's. The code was tested in two NVIDIA 295 GTX cards, each card has two GPU's inside. Due to the data access patterns and the limited amount of the shared memory available, we use texture memory to load data. Note that the texture memory can only be read inside the kernels and updated between kernels. The lattice array is a \mbox{\texttt{float4}} type array and the four sites of the \mbox{\texttt{float4}} are used to store the $a_0$, $a_1$, $a_2$ and $a_3$. All the C/CUDA operators (plus, multiplication, trace, determinant, and so on) was developed using the SU(2) generators.

The CUDA code implementation consists of 7 kernels, one kernel to generate and store random numbers, this part is based on \cite{random}, two kernels to the type of the initialization of the lattice, cold start or hot start, a kernel to perform the heat bath method, a kernel to compute the plaquette at each lattice site, a kernel to compute the Polyakov loop at each lattice site and a kernel to perform the sum over all the lattice sites for the plaquette and/or Polyakov loop.

\begin{figure}[h]
\begin{centering}
    \subfloat[\label{Grid}]{
\begin{centering}
    \includegraphics[width=4.0cm]{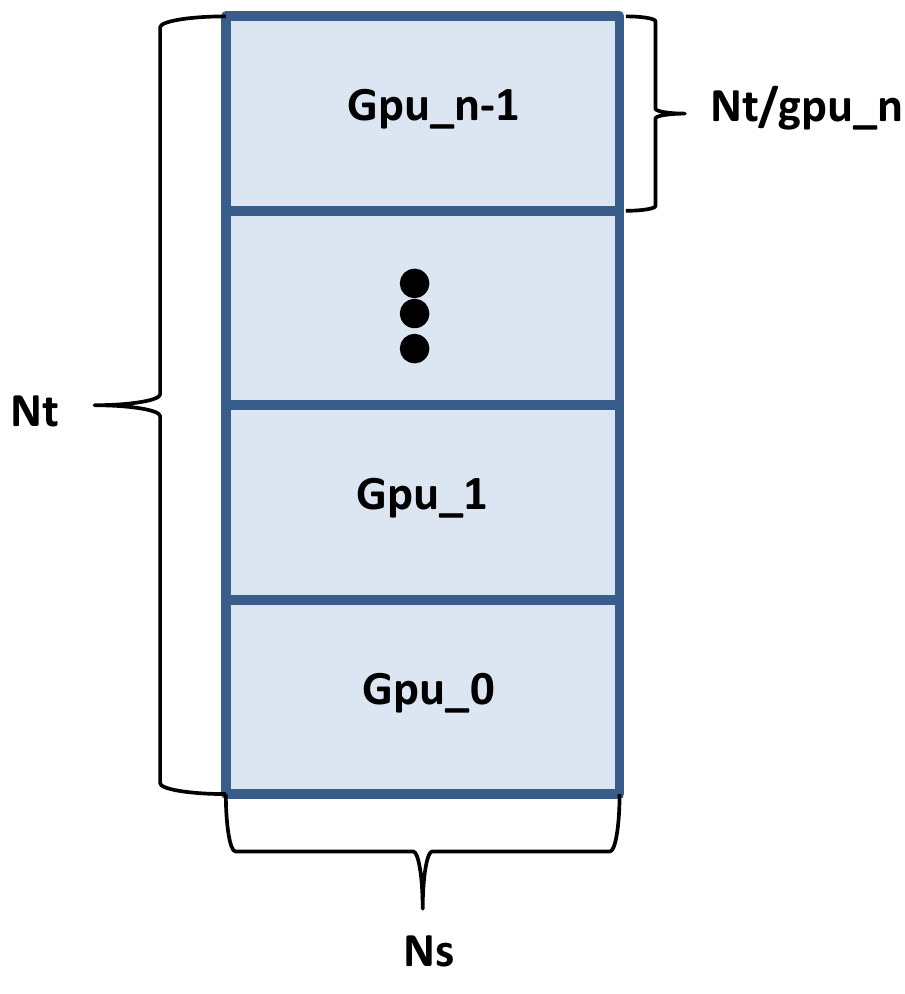}
\par\end{centering}}
    \subfloat[\label{Grid2}]{
\begin{centering}
    \includegraphics[width=4cm]{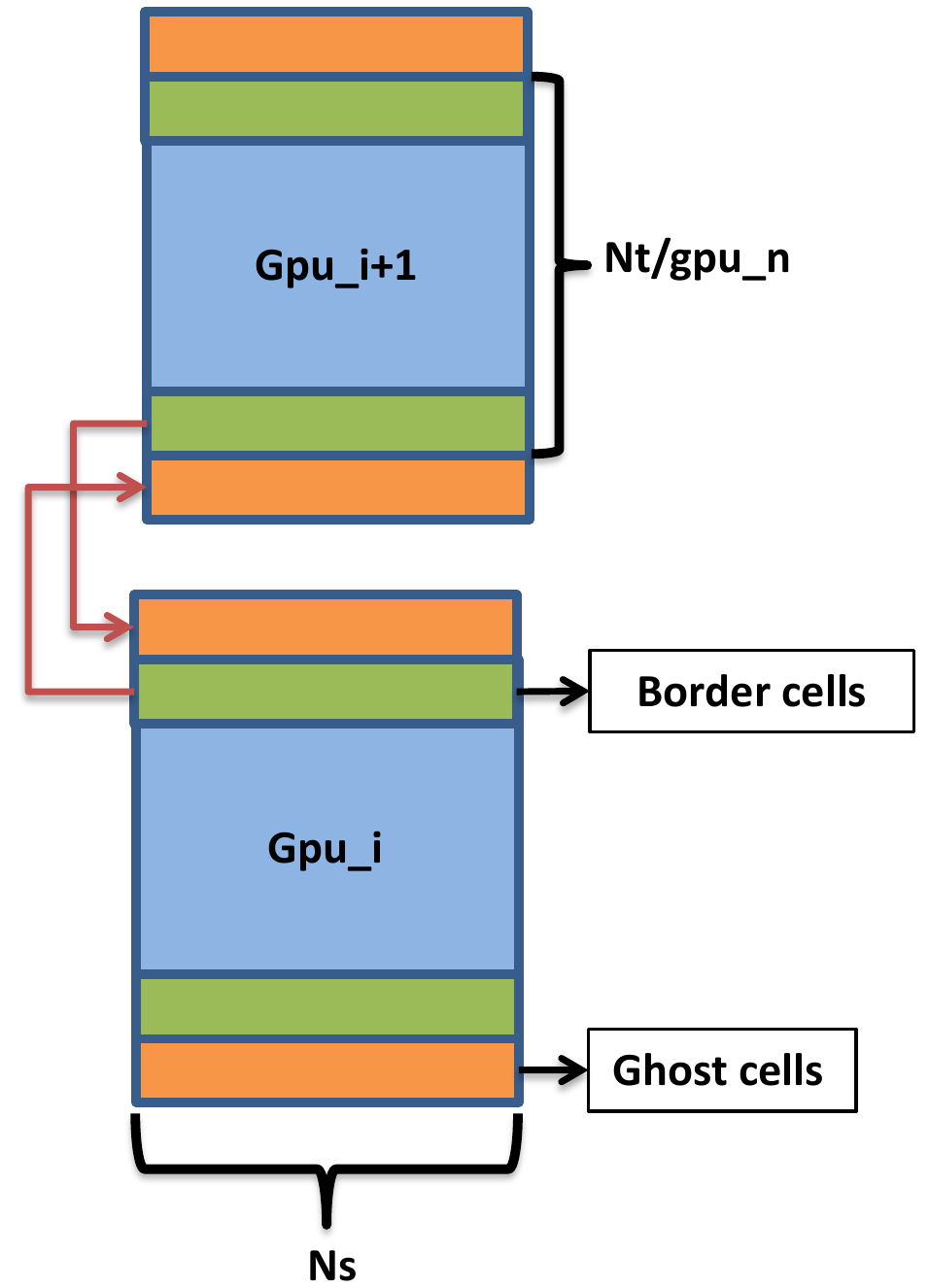}
\par\end{centering}}
\par\end{centering}
    \caption{Schematic view of the lattice size part in each GPU.}
    \label{openmp_grid}
\end{figure}

The Multi-GPU part was implemented using CUDA and OPENMP, each CPU thread controls one GPU. Each GPU computes $Ns \times \frac{Nt}{num.\, gpu's}$, the total length of the array in each GPU is $Ns\times (\frac{Nt}{num.\, gpu's} + 2)$, see figure \ref{openmp_grid}. At each iteration, the links are calculated separately by even and odd sites and direction, $\mu$, before calculate the following direction, the border cells need to be exchanged between each GPU. In order to exchange border cells between GPU's it is necessary to copy these cells to CPU memory and then synchronize each CPU thread with the command \mbox{\texttt{\#pragma omp barrier}} before updating the GPU memory, ghost cells. 

\subsection{GPU Performance}

\begin{figure}[h]
\begin{centering}
    \subfloat[Execution time.\label{Speed_mGPU}]{
\begin{centering}
    \includegraphics[width=6cm]{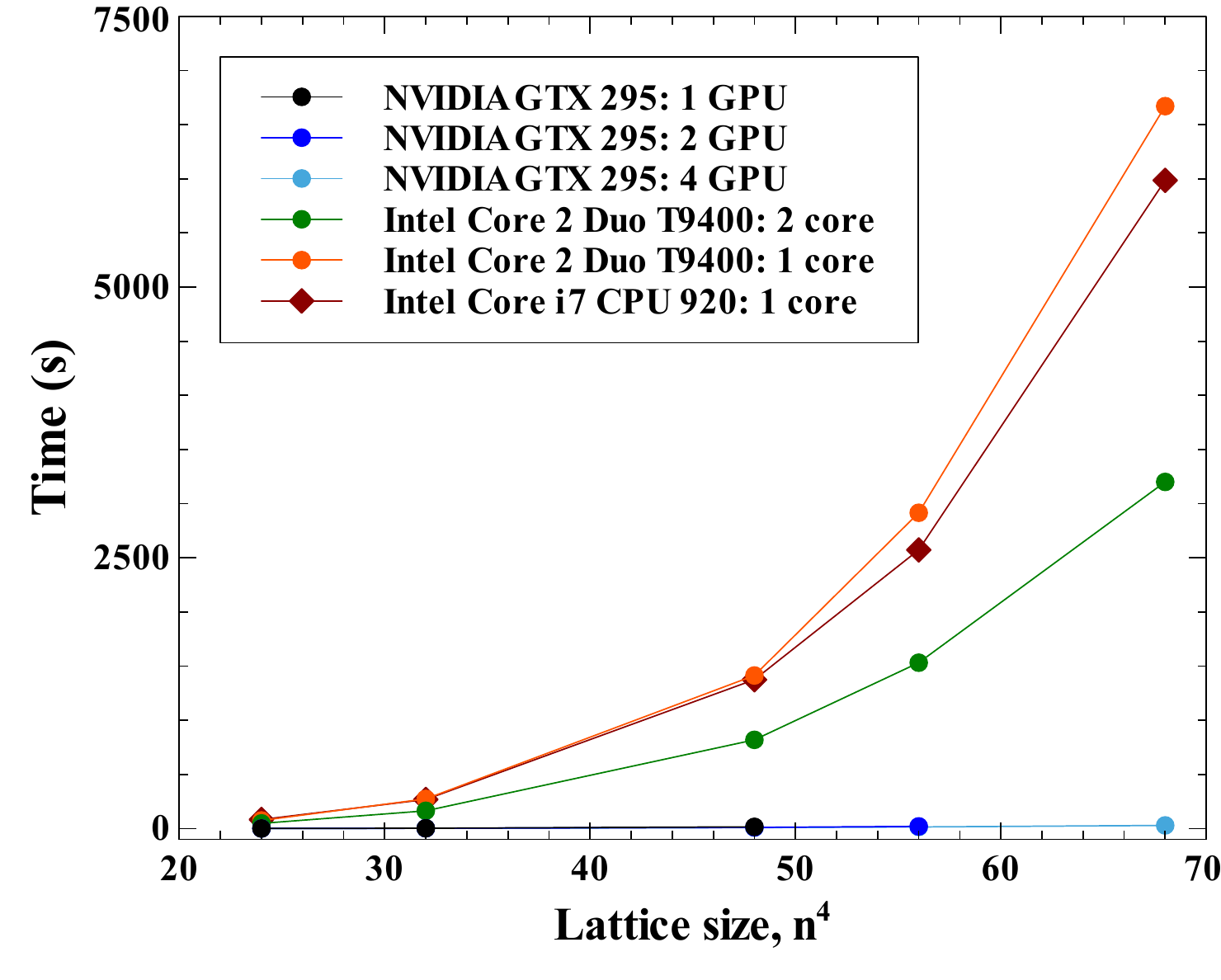}
\par\end{centering}}
    \subfloat[Speed over Intel Core i7 CPU 920 with 1 core\label{speed_0}]{
\begin{centering}
    \includegraphics[width=9cm]{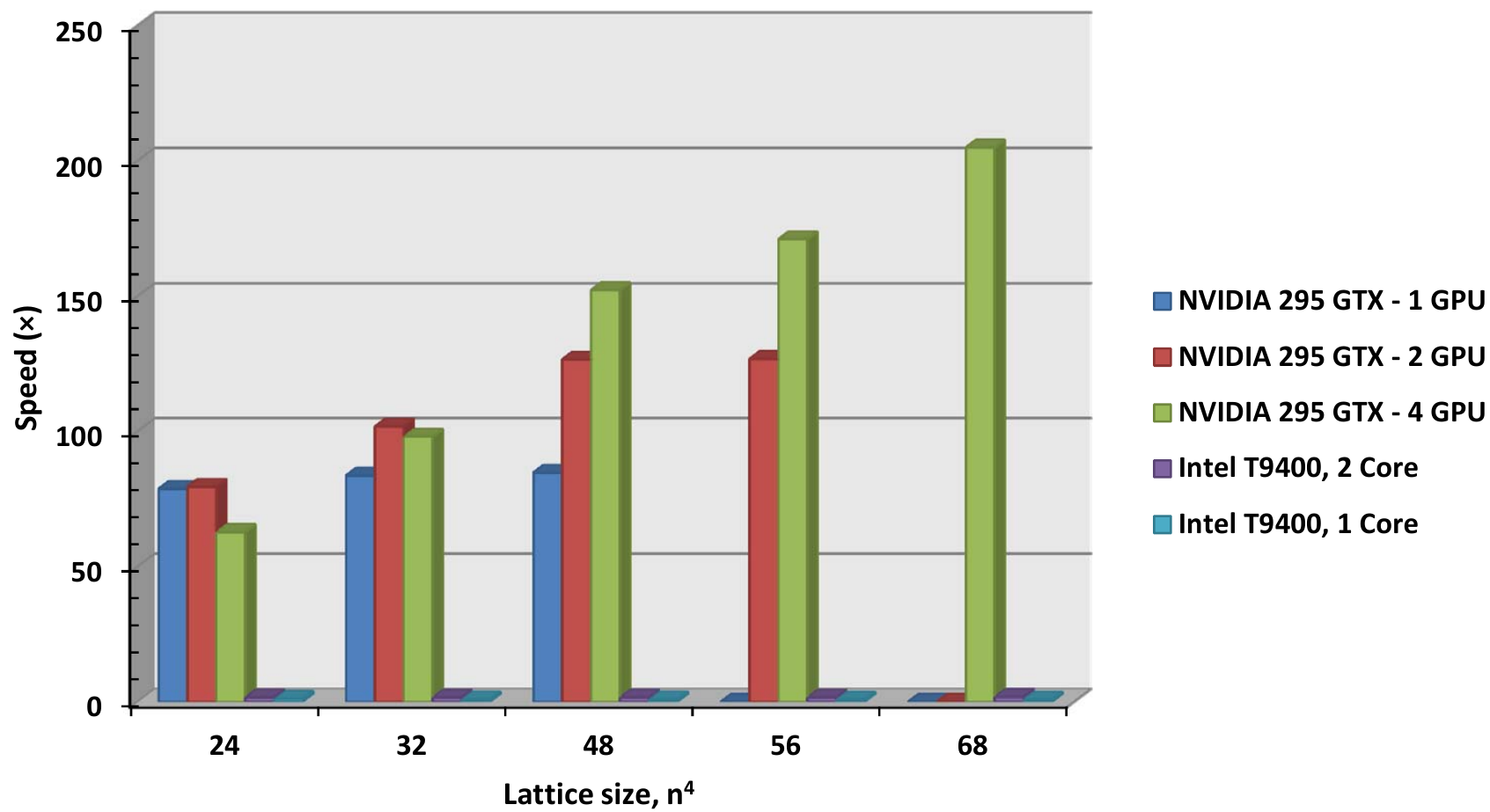}
\par\end{centering}}
\par\end{centering}
    \caption{Execution time, \subref{Speed_mGPU}, and speed performance over one core CPU, \subref{speed_0}, for different lattice sizes at $\beta = 6.0$, hot start initialization of the lattice, 100 iterations of the heat bath method and the calculation of the mean average plaquette at each iteration }
    \label{performance}
\end{figure}

\begin{figure}[h]
\begin{centering}
    \subfloat[Summary Plot with a single GPU for a $32^4$ lattice\label{sum_plot_32_4_6_100i_tex}]{
\begin{centering}
    \includegraphics[width=6.3cm]{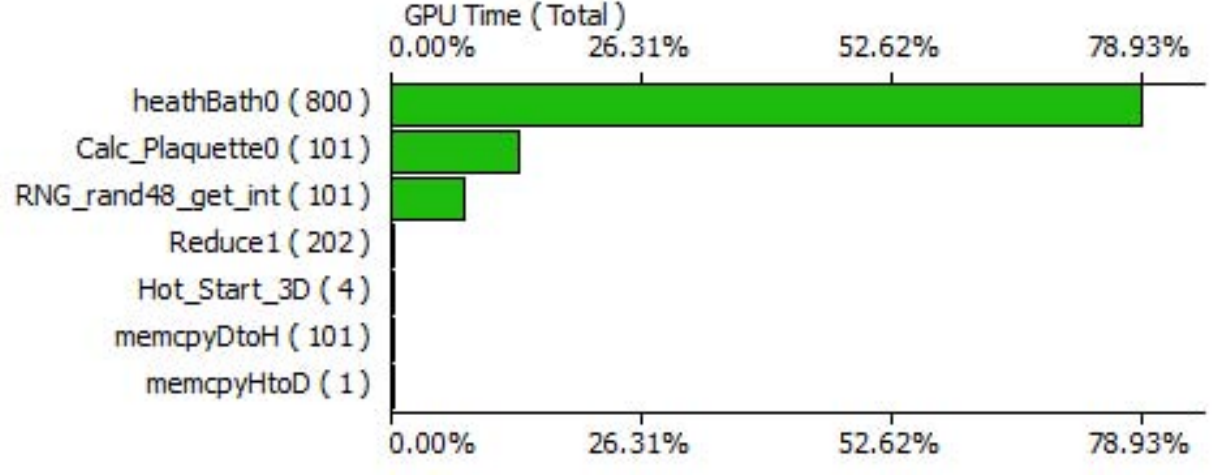}
\par\end{centering}}
    \subfloat[Summary Plot with 4 GPU's for a $56^4$ lattice\label{sum_plot_56_6_100i_tex_4gpu}]{
\begin{centering}
    \includegraphics[width=8.3cm]{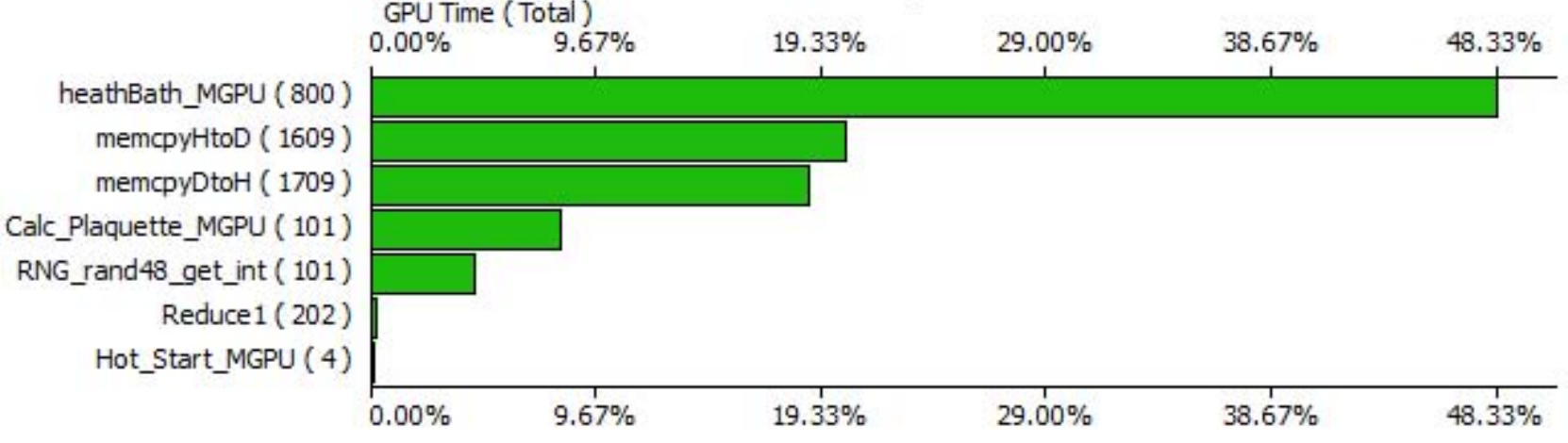}
\par\end{centering}}
\par\end{centering}
    \caption{Performance with NVIDIA Visual Profiler. memcpyHtoD means memory transfers from Host (CPU) to Device (GPU) memory; memcpyDtoH means memory transfers from Device to Host memory; Hot\_Start* is the kernel to initialize the link of the lattice site with a random SU(2) matrix; heatbath* is the kernel for the heat bath method (the total number is given by the number of iterations (100) times the number of directions (4) times two (first the even then the odd sites are computed which gives 800); Calc\_Plaquette* is the kernel for compute the plaquette for every lattice site; Reduce1 is the kernel to perform the sum of the plaquette in the entire lattice; RNG\_rand48\_get\_int is the kernel to generate random numbers}
    \label{per_visual_profiler}
\end{figure}

In order to test the GPU performance, we measure the execution time for the CUDA code implementation in one, two and four GPU's and the serial code in CPU (Intel Core i7 CPU 920 and Mobile DualCore Intel Core 2 Duo T9400) for different lattice sizes at $\beta = 6.0$, hot start initialization of the lattice, 100 iterations of the heat bath method and the calculation of the mean average plaquette at each iteration, see figure \ref{performance}. In figure \ref{Speed_mGPU}, we show the execution time and in figure \ref{speed_0} the speed performance with the Intel Core i7 CPU 920 with only one core. With NVIDIA Visual Profiler it is possible to analyze the percentage taken by each kernel of the code, see figure \ref{per_visual_profiler}.

\subsection{Mean Average Plaquette and Polyakov Loop}

\begin{figure}[h]
\begin{centering}
    \subfloat[Lattice size $8^4$\label{beta_iter_8_cold_4d}]{
\begin{centering}
    \includegraphics[width=6.7cm]{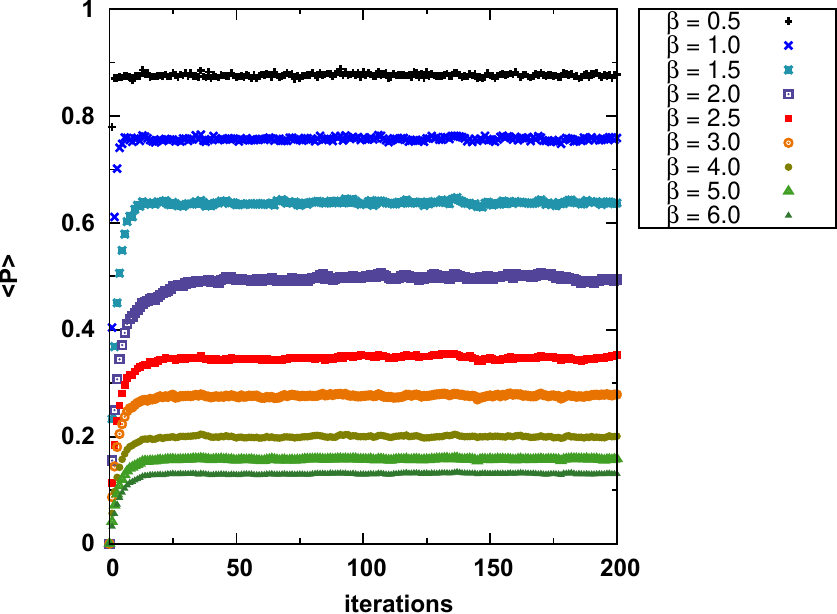}
\par\end{centering}}
    \subfloat[Lattice size $48^4$\label{beta_iter_48_cold_4d}]{
\begin{centering}
    \includegraphics[width=6.7cm]{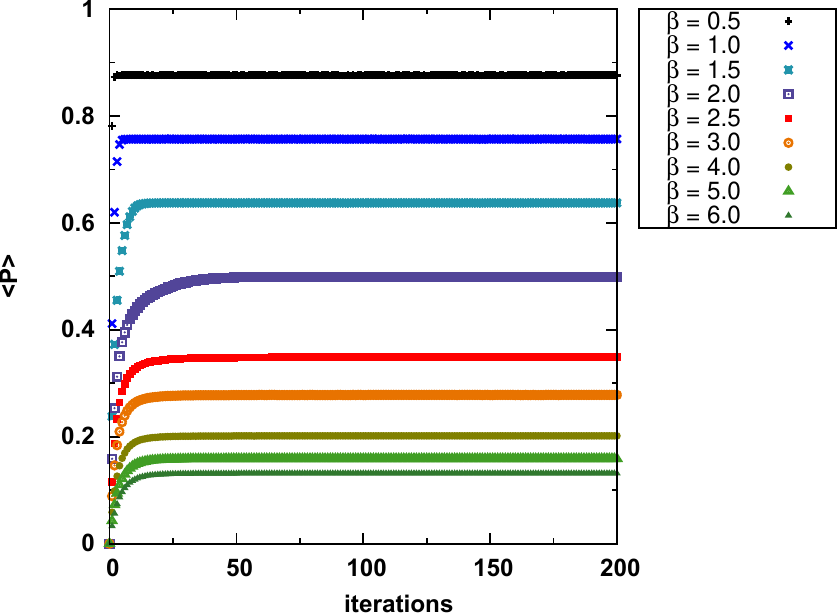}
\par\end{centering}}
\par\end{centering}
    \caption{Mean plaquette per iteration for different $\beta$ and two lattice sizes, $8^4$ and $48^4$, with cold start initialization and 200 iterations of the heat bath method.}
    \label{mean_plaq_iter}
\end{figure}

\begin{figure}[h]
\begin{centering}
    \includegraphics[width=7.5cm]{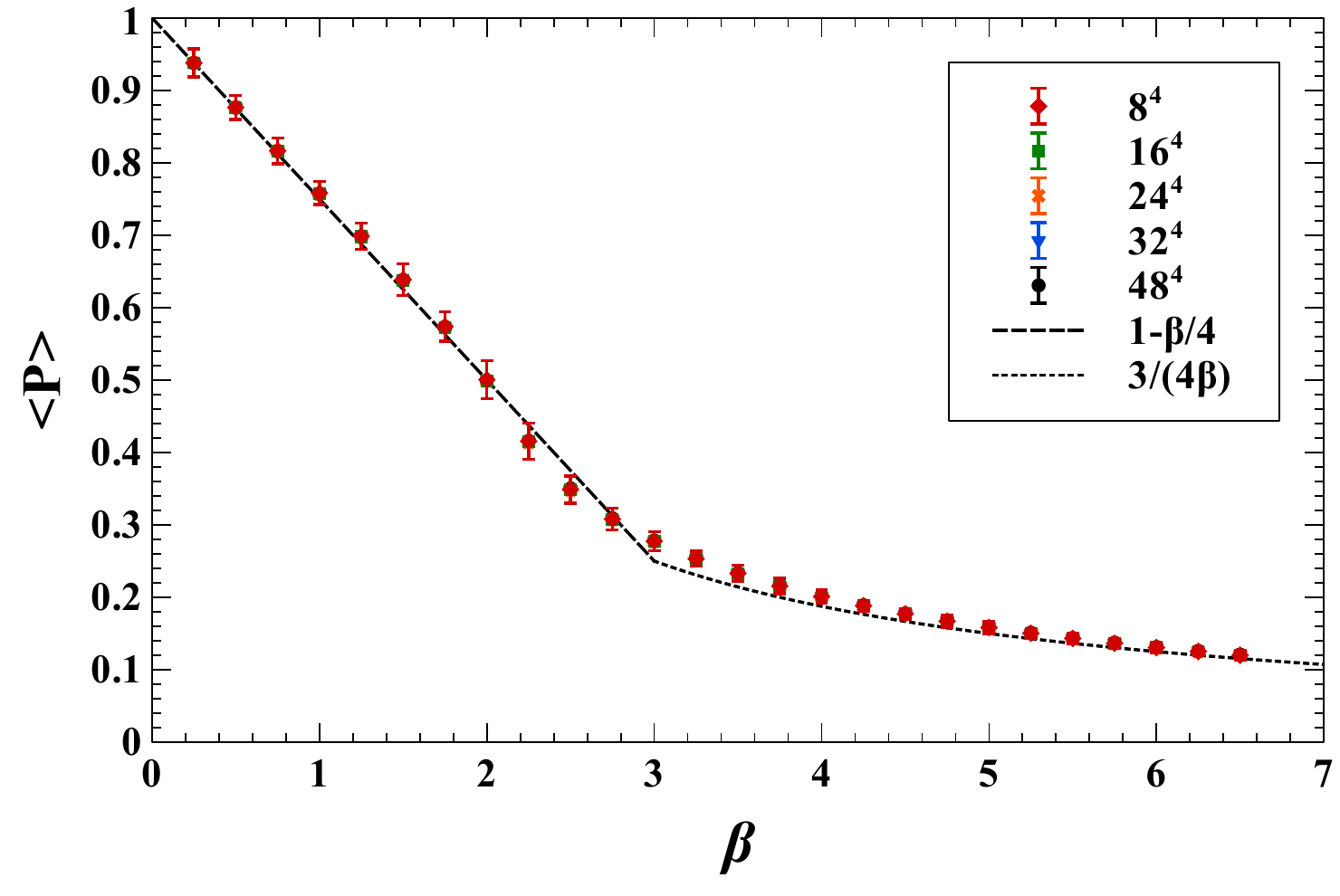}
\par\end{centering}
    \caption{$\beta$ dependence of the average plaquette from Monte Carlo simulation for several lattice sizes (data points) and analytic predictions (denoted by dashed lines).}
    \label{mean_avg_plaquette}
\end{figure}

In this section we present the mean average plaquette at each heat bath iteration, figure \ref{mean_plaq_iter}, and the mean average plaquette for each $\beta$ for 5000 configurations, \ref{mean_avg_plaquette} with different lattice sizes.

\begin{figure}[h]
\begin{centering}
    \subfloat[Mean average Polyakov loop versus $\beta$. $\beta$ dependence of the average Polyakov loop from Monte Carlo simulation for different temporal lattice sizes.\label{polyakov_loop_48_L}]{
\begin{centering}
   \includegraphics[width=7.6cm]{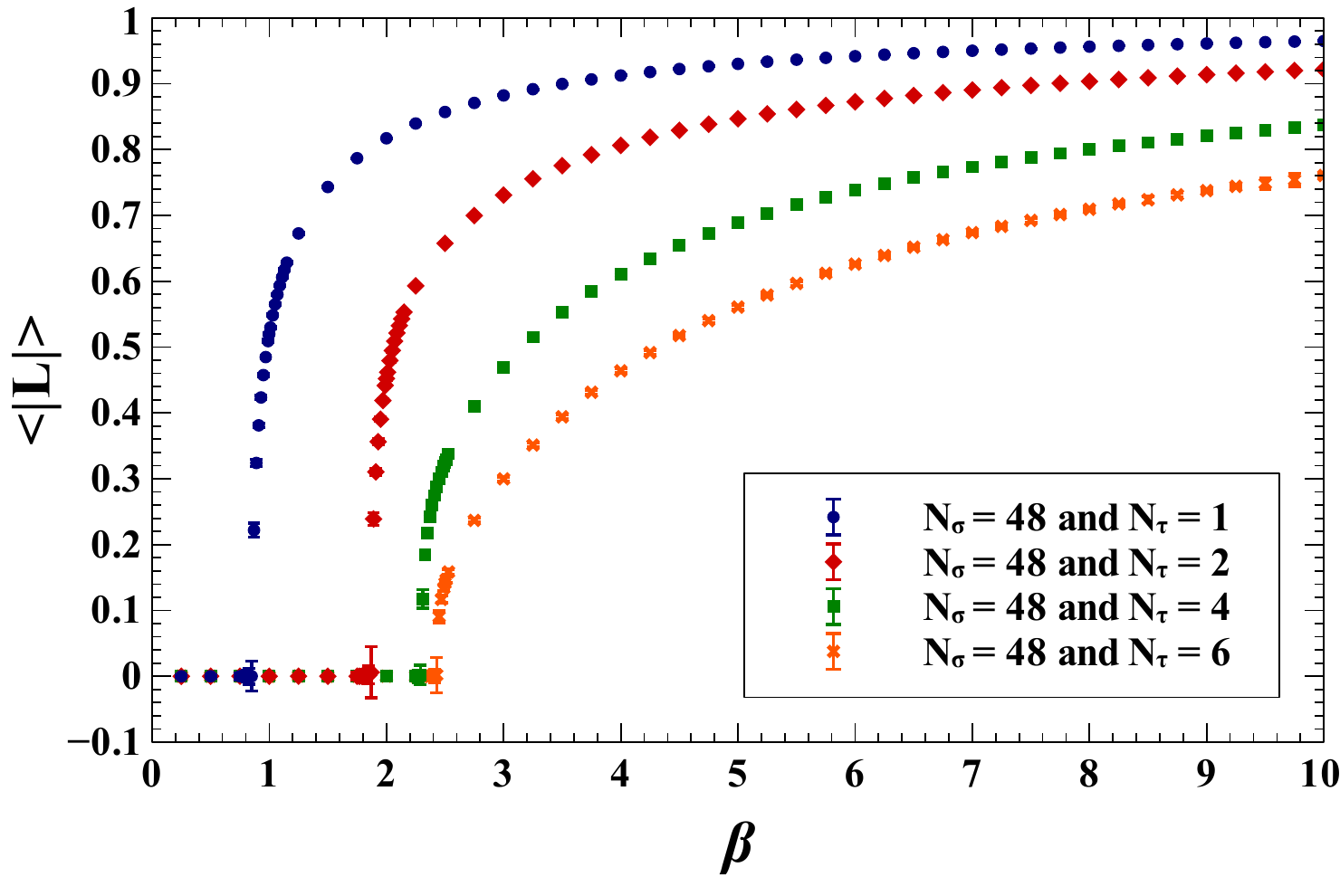}
\par\end{centering}}
    \subfloat[Mean average Polyakov loop versus $T/T_c$.\label{PL_TTc}]{
\begin{centering}
    \includegraphics[width=6.4cm]{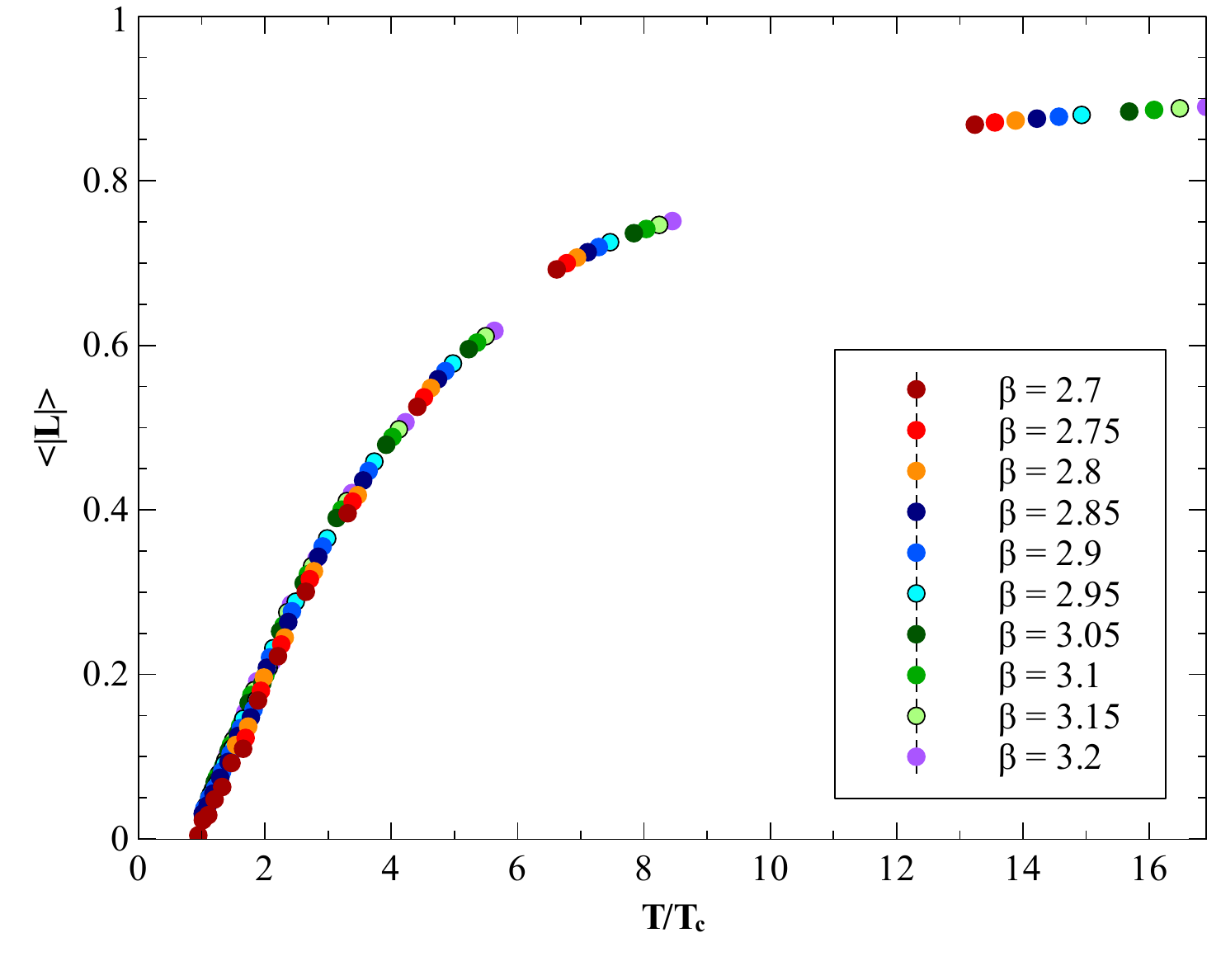}
\par\end{centering}}
\par\end{centering}
    \caption{Mean average Polyakov loop}
    \label{mean_avg_polyakov loop}
\end{figure}

In figure \ref{mean_avg_polyakov loop} we present the mean average Polyakov loop for several $\beta$ and temporal lattice sizes, \ref{polyakov_loop_48_L}, and the mean average Polyakov loop versus the temperature ($T/T_c$), \ref{PL_TTc}, for 200 000 heat bath iterations.

\section{Conclusions}
With 2 NVIDIA GTX 295, we were able to obtain more than $200\times$ the performance over the Intel Core i7 CPU 920 with 1 core for a $68^4$ lattice using single precision.

It's not possible to generate SU(2) configurations using only GPU shared memory due to the limited amount of shared memory available.
The limited number of registers also affects the GPU performance.
Using texture memory in this problem, we were able to achieve the high performance.

The occupancy and performance of the GPU's is strongly linked to the number of threads per block, registers per thread, shared memory per block, memory access, read and writing, patterns. To maximize performance it is necessary to ensure that the memory access is coalesced and to minimize copies between GPU and CPU memories.

Using texture memory there is no problems with coalesced memory, it cannot be updated inside kernels but only updated between kernels, it is very useful for complicated problems where memory access (read/write) cannot be achieved in a coalesced pattern. However, the hardware doesn't support double precision as texture format, it is possible to use \mbox{\texttt{int2}} and \mbox{\texttt{\_\_hiloint2double}} and cast it to double. This was not implemented in the code.

NVIDIA Visual Profiler is a useful tool to help measure execution time, performance, number of registers and shared memory used by each kernel and other functions.

Although NVIDIA 295 GTX supports double precision, the performance is much slower because it has only one double precision unit for each multiprocessor. We expect that the new generation of the NVIDIA cards, the Fermi architecture, is better for double precision calculations with $8\times$ the peak double precision floating point performance over previous ones.

\acknowledgments
This work was financed by the FCT contracts POCI/FP/81933/2007, CERN/FP/83582/2008, PTDC/FIS/100968/2008 and CERN/FP/109327/2009.
We thank Marco Cardoso and Orlando Oliveira for useful discussions.

\end{document}